\documentclass[a4paper, 12pt]{article}
\usepackage[english]{babel}
\usepackage[utf8]{inputenc}
\usepackage[T1]{fontenc}
\usepackage{amsmath}
\usepackage{graphicx}
\usepackage[colorinlistoftodos]{todonotes}
\usepackage[a4paper,left=2.5cm,right=2.5cm,top=2.cm,bottom=2cm]{geometry}
\usepackage{authblk}
\usepackage{txfonts}
\usepackage{bbold} 
\usepackage{amsopn} 
\usepackage{graphicx}
\newtheorem{example}{Example}
\newtheorem{theorem}{Theorem}
\newtheorem{proposition}{Proposition}
\newtheorem{definition}{Definition}

\newenvironment{proof}{\noindent\textit{Proof~~}}
{\nolinebreak[4]\hfill$\blacksquare$\\\par}

\title{Nonclassical rules in quantum games}

\author{Piotr Fr\k{a}ckiewicz}
\date{%
  Institute of Exact and Technical Sciences\\ Pomeranian University in Słupsk, Poland
}

\begin{document}
\maketitle

\begin{abstract} 
Over the last twenty years of research on quantum game theory have given us many ideas of how quantum games could be played. One of the most prominent ideas in the field is a model of quantum playing a $2\times 2$ game introduced by J. Eisert, M. Wilkens and M. Lewenstein. The scheme assumes that players' strategies are unitary operations the players act on the maximally entangled two-qubit state. The quantum nature of the scheme has been under discussion since the article by Eisert et al. came out. The aim of our paper is to identify some of non-classical features of the quantum scheme. 
\end{abstract}
\section{Introduction}
The scheme defined by J. Eisert, M. Wilkens and M. Lewenstein \cite{eisert} was one of the first formal protocols of playing quantum game, and  is definitely one of the most used schemes for quantum games. This conclusion is confirmed by the number of citations of the article (around 500 citations according to Web of Knowledge).   The scheme generalizes a $2\times 2$ game in the sense that the game generated by the Eisert-Wilkens-Lewenstein (EWL) scheme with unitary strategies restricted to some type of one-parameter operators is equivalent to the classical game. The seminal paper \cite{eisert} and the subsequent papers \cite{duenhanced}-\cite{unawarenessfracor} are just a very smart part of the huge literature devoted to the EWL scheme. It was shown in \cite{eisert} that a quantum way of playing the Prisoner's Dilemma game can lead to a reasonable and Pareto efficient outcome. Further research has shown, for example, that players can benefit from the use of quantum strategies in symmetric $2\times 2$ games \cite{flitneygeneralized}. The Eisert-Wilkens-Lewenstein (EWL) scheme can also be extended to consider extensive-form games \cite{fracorextensive}. It was also shown that the EWL scheme can be implemented with a quantum computer \cite{prevedel}, \cite{nebula}. 

Despite the significance of the scheme in the development of quantum game theory, doubts arise as to quantum nature of the EWL game. These concerns include the following:
\begin{itemize}
\item does the quantum solution provided by the EWL scheme really solve the input classical game?
\item can the quantum solution be obtained in a classical game?
\end{itemize}
These questions were raised in \cite{enk}. By considering the Prisoner's Dilemma game the authors come to the conclusion that the EWL scheme does not imply a quantum mechanical game. Moreover, according to \cite{enk}, the solution (Nash equilibrium) resulting from playing the EWL game does not appear to solve the original game. 

Recently, there have been discussions about van Enk and Pike's arguments.  It is claimed in \cite{Vyas} that the EWL approach to the Hawk-Dove game 
 enables the players to obtain a game result that is not achievable in the classical game. As a result, it was concluded in \cite{Vyas} that a quantum game cannot be fully modeled by the classical game. Shortly after appearing  \cite{Vyas}, B. Groisman \cite{groisman} suggested that the scheme used by N. Vyas and C. Benjamin changes the rules of the original game. Hence, the author stated that the solution provided in  \cite{Vyas} cannot be treated as a quantum extension of the classical game. 
 
In light of the above, it can be seen that the problem of quantumness of the EWL scheme is not resolved. The purpose of this article is, on the one hand, to show that the form of the scheme considered in \cite{enk}, \cite{Vyas} and \cite{groisman} does not fully describe the EWL scheme, on the other hand, to draw attention to another non-classical properties of the scheme. 

\section{Preliminaries on game theory}
This section is based on \cite{maszler} and \cite{sungjuang}. We review relevant material connected with the notion of strategic-form games and payoff regions in that games.

The basic model of games studied in game theory is a game in strategic form.
\begin{definition}\textup{\cite{maszler}}~
A game in strategic form (or in normal form) is an ordered triple 
\begin{equation}\label{strgame}
(N, (S_{i})_{i\in N}, (u_{i})_{i\in N}), 
\end{equation}
in which
\begin{itemize}
\item $N = \{1,2, \dots, n\}$ is a finite set of players,
\item $S_{i}$ is the set of strategies of player $i$, for every player $i\in N$,
\item $u_{i}\colon S_{1}\times S_{2} \times \cdots \times S_{n} \to \mathbb{R}$ is a function associating each vector of strategies $s= (s_{i})_{i\in N}$ with the payoff $u_{i}(s)$ to player $i$, for every player $i\in N$.
\end{itemize}
\end{definition}
\noindent In the case of a finite two-person game, i.e., $N=\{1,2\}$, $S_{1}= \{0,1, \dots, m-1\}$, $S_{2} = \{0,1, \dots, n-1\}$, the game can be written as a bimatrix with entries $(u_{1}(s), u_{2}(s))$, 
\begin{equation}\label{bimatrix}
\bordermatrix{& 0 & 1 & \cdots & n-1 \cr
0 & (a_{00}, b_{00}) & (a_{01}, b_{01}) & \cdots  & (a_{0,n-1}, b_{0,n-1}) \cr 
1 & (a_{10}, b_{10}) & (a_{11}, b_{11}) & \cdots  & (a_{1,n-1}, b_{1,n-1}) \cr 
\vdots & \vdots & \vdots & \ddots & \vdots \cr 
m-1 & (a_{m-1,0}, b_{m-1,0}) & (a_{m-1,1}, b_{m-1,1}) & \cdots & (a_{m-1, n-1}, b_{m-1, n-1})
}.
\end{equation}

The elements of $S_{i}$ are called the pure strategies of player $i$. The set of pure strategy vectors (profiles) is $\prod^n_{i=1}S_{i}$. A mixed strategy of player $i$ is a probability distribution over $S_{i}$. We denote the set of mixed strategies of player $i$ by $\Delta(S_{i})$. The set of mixed strategy profiles is $\prod^n_{i=1}\Delta(S_{i})$. In particular, if $S_{i} = \{s^i_{0}, s^i_{1}\}$, player $i$'s set of mixed strategies will be denoted by 
\begin{equation}
\{[p_{1}(s^i_{0}), p_{2}(s^i_{1})] \colon p_{1} \geq 0, p_{2} \geq 0, p_{1} + p_{2} = 1\}.
\end{equation}
A correlated strategy is a probability distribution over $\prod^n_{i=1}S_{i}$. The set of correlated strategies is denoted by $\Delta(\prod^n_{i=1}S_{i})$. 

Let $u_{i}\colon \prod^n_{i=1}S_{i} \to \mathbb{R}$ be the payoff function of player $i$ in $(N, (S_{i})_{i\in N}, (u_{i})_{i\in N})$. Then the payoff functions $u_{i}\colon \prod^n_{i=1}\Delta(S_{i}) \to \mathbb{R}$ and $u_{i}\colon \Delta(\prod^n_{i=1}S_{i}) \to \mathbb{R}$ are defined by the expected values of $u_{i}\colon \prod^n_{i=1}S_{i} \to \mathbb{R}$ determined by mixed strategies $\sigma \in \prod^n_{i=1}\Delta(S_{i})$ and probability distributions over $\prod^n_{i=1}S_{i}$, respectively. Let us define the vector-valued payoff function $u\colon \prod^n_{i=1}S_{i} \to \mathbb{R}^n$ by $u(s) = (u_{1}(s), \dots, u_{n}(s))$, $s \in \prod^n_{i=1}S_{i}$.
\begin{definition}\textup{\cite{sungjuang}} Let $(N, (S_{i})_{i\in N}, (u_{i})_{i\in N})$ be a finite strategic-form game. The ranges 
\begin{equation}
R_{\textup{pu}} = u\left(\prod^n_{i=1}S_{i}\right), \quad R_{\textup{nc}} = u\left(\prod^n_{i=1}\Delta(S_{i})\right), \quad R_{\textup{co}} = u\left(\Delta\left(\prod^n_{i=1}S_{i}\right)\right)
\end{equation}
are called the pure-payoff region, the noncooperative payoff region and the cooperative payoff region, respectively. \label{defR}
\end{definition}
The notion of Nash equilibrium is one of the most important solution concepts in noncooperative game theory. It defines a strategy vector at which each strategy is a best reply to the strategies of the other players. 
\begin{definition}\label{Nash} \textup{\cite{maszler}}~
A strategy vector $s^*=(s^*_{1}, s^*_{2}, \dots, s^*_{r})$ is a Nash equilibrium if for each player $i\in N$ and each strategy $s_{i} \in S_{i}$ the following is satisfied:
\begin{equation}\label{Nashcondition}
u_{i}(s^*) \geq u_{i}(s_{i}, s^*_{-i}),
\end{equation}
where $s^*_{-i} = (s^*_{1}, \dots, s^*_{i-1}, s^*_{i+1}, \dots, s^*_{r})$.
\end{definition}
In particular, if a strategic form game is described in bimatrix form, Nash equilibrium can be defined as follows:
\begin{definition}
A position $(i,j)$ in a bimatrix game (\ref{bimatrix}) is a Nash equilibrium if 
\begin{equation}
    a_{ij} \geq a_{kj} ~~\text{for all}~~ k\in \{0,1, \dots, m-1\} 
\end{equation}
and
\begin{equation}
    b_{ij} \geq a_{il} ~~\text{for all}~~ l\in \{0,1, \dots, n-1\}.
\end{equation}
\end{definition}
\section{The Eisert-Wilkens-Lewenstein scheme}
The Eisert-Wilkens-Lewenstein (EWL) scheme is a model of a normal-form framework. It concerns bimatrix $2\times 2$ games -- two person strategic form games with two-element sets of strategies that can be written as  
\begin{equation}\label{bimatrix}
\bordermatrix{& s^2_{0} & s^2_{1} \cr
s^1_{0} & (a_{00}, b_{00}) & (a_{01}, b_{01}) \cr 
s^1_{1} & (a_{10}, b_{10}) & (a_{11}, b_{11}) }.
\end{equation}
 In the EWL scheme, players' strategies are unitary operators that each of two players acts on a maximally entangled quantum state. In the literature there are a few descriptions of the EWL scheme that are strategically equivalent. In what follows, we recall the general $n$-person scheme we adapted for the purpose of our research.

\begin{definition}\textup{\cite{strongfracor}}\label{ewldefinition}
Let us consider a strategic game $\Gamma = (N, (S_{i})_{i\in N}, (u_{i})_{i\in N})$ with $S_{i} = \{s^i_{0}, s^i_{1}\}$ for each $i\in N$. The Eisert-Wilkens-Lewenstein approach to game $\Gamma$ is defined by a triple $\Gamma_{EWL} = (N, (D_{i})_{i\in N}, (v_{i})_{i\in N})$, where 
\begin{itemize}
\item $N = \{1,2, \dots n\}$ is the set of players.
\item $D_{i}$ is a set of unitary operators from $\mathsf{SU}(2)$. A possible parametrization of $U \in \mathsf{SU}(2)$ is
\begin{equation}\label{2parameter}
U_{i}(\theta_{i}, \alpha_{i}, \beta_{i}) = \begin{pmatrix} 
\mathrm{e}^{\mathrm{i}\alpha_{i}}\cos\frac{\theta_{i}}{2} & \mathrm{i}\mathrm{e}^{\mathrm{i}\beta_{i}}\sin\frac{\theta_{i}}{2}\\
\mathrm{i}\mathrm{e}^{-\mathrm{i}\beta_{i}}\sin\frac{\theta_{i}}{2} & \mathrm{e}^{-\mathrm{i}\alpha_{i}}\cos\frac{\theta_{i}}{2}
\end{pmatrix}, \quad \theta_{i} \in [0,\pi], \alpha_{i}, \beta_{i} \in [0,2\pi).
\end{equation}
\item $v_{i}\colon D_{1}\otimes D_{2} \otimes \dots \otimes D_{n} \to \mathbb{R}$ is a payoff function given by
\begin{equation}\label{ewlpayofffunction}
v_{i}\left(\bigotimes^n_{i=1} U_{i}(\theta_{i}, \alpha_{i}, \beta_{i})\right) = \operatorname{tr}(|\Psi\rangle \langle \Psi| M_{i}),
\end{equation}
where
\begin{equation}
\begin{split}\label{finalstate7}
&|\Psi\rangle = J^{\dag}\left(\bigotimes^n_{i=1} U_{i}(\theta_{i}, \alpha_{i}, \beta_{i})\right) J|0\rangle^{\otimes n}, \quad J = (\mathbb{1}^{\otimes n} + \mathrm{i}\sigma_{x}^{\otimes n})/\sqrt{2},\\
&M_{i} = \sum_{j_{1}, \dots, j_{n}\in \{0,1\}}a^i_{j_{1},\dots, j_{n}}|j_{1},\dots, j_{n}\rangle \langle j_{1}, \dots, j_{n}|, 
\end{split}
\end{equation}
and $a^i_{j_{1},\dots, j_{n}} \in \mathbb{R}$ are payoffs of player $i$ in $\Gamma$ given by equation $a^i_{j_{1},\dots, j_{n}} = u_{i}(s^1_{j_{1}}, \dots, s^n_{j_{n}})$.
\end{itemize}
\end{definition}
In particular, the EWL approach to a $2\times 2$ game (\ref{bimatrix}) results in the following vector-valued payoff functions:
\begin{multline}\label{payoffvector2x2}
v(U_{1}(\theta_{1}, \alpha_{1}, \beta_{1}), U_{2}(\theta_{2}, \alpha_{2}, \beta_{2})) = (a_{00}, b_{00})\left(\cos{(\alpha_{1} + \alpha_{2})}\cos{\frac{\theta_{1}}{2}}\cos{\frac{\theta_{2}}{2}} + \sin{(\beta_{1} + \beta_{2})\sin{\frac{\theta_{1}}{2}}\sin{\frac{\theta_{2}}{2}}}\right)^2 \\ 
+ (a_{01}, b_{01})\left(\sin{(\alpha_{2}-\beta_{1})}\sin{\frac{\theta_{1}}{2}}\cos{\frac{\theta_{2}}{2}} + \cos{(\alpha_{1}-\beta_{2})}\cos{\frac{\theta_{1}}{2}}\sin{\frac{\theta_{2}}{2}}\right)^2\\
+ (a_{10}, b_{10})\left(\cos{(\alpha_{2}-\beta_{1})}\sin{\frac{\theta_{1}}{2}}\cos{\frac{\theta_{2}}{2}} + \sin{(\alpha_{1}-\beta_{2})}\cos{\frac{\theta_{1}}{2}}\sin{\frac{\theta_{2}}{2}}\right)^2 \\ 
+ (a_{11}, b_{11})\left(\cos{(\beta_{1}+\beta_{2})}\sin{\frac{\theta_{1}}{2}}\sin{\frac{\theta_{2}}{2}} - \sin{(\alpha_{1} + \alpha_{2})}\cos{\frac{\theta_{1}}{2}}\cos{\frac{\theta_{2}}{2}}\right)^2.
\end{multline}
\section{Problem of classical strategies in the EWL scheme}\label{mainsec}
The EWL scheme constitutes a generalization of the classical way of playing the game. It is known that the EWL game becomes equivalent to the classical one by restricting the unitary strategy sets of the players. In the case of a bimatrix game (\ref{bimatrix}), the scheme 
\begin{equation}\label{ewlclassic}
\Gamma_{EWL} = (\{1,2\}, (D_{i})_{i\in \{1,2\}}, (v_{i})_{i\in \{1,2\}})
\end{equation} 
is equivalent to (\ref{bimatrix}) if 
\begin{equation}\label{oneparameterd1}
D_{1} = D_{2} = \{U(\theta,0,0) \mid \theta \in [0,\pi]\}. 
\end{equation}
If the players choose $U_{1}(2\arccos\sqrt{p}, 0,0), U_{2}(2\arccos\sqrt{q}, 0,0) \in \{U(\theta,0,0) \mid \theta \in [0,\pi]\}$ then the resulting payoff vector is of the form
\begin{multline}\label{classicalpayoffmixed}
v(U_{1}(\theta_{1}, \alpha_{1}, \beta_{1}), U_{2}(\theta_{2}, \alpha_{2}, \beta_{2}))\\
= (a_{00}, b_{00})pq + (a_{01}, b_{01})p(1-q) + (a_{10}, b_{10})(1-p)q + (a_{11}, b_{11})(1-p)(1-q).
\end{multline}
This is the same as the payoff vector corresponding to a profile of classical mixed strategies 
\begin{equation}
\left([p(s^1_{0}), (1-p)(s^1_{1})],  [q(s^2_{0}), (1-q)(s^2_{1})]\right).
\end{equation}
On the other hand, player 1 and player 2's classical mixed strategies in the EWL scheme can also be modeled by quantum operations
\begin{equation}\label{cmstrategies}
\mathcal{C}_{p}(\rho) = p\mathbb{1}\rho \mathbb{1} + (1-p)U(\pi,0,0)\rho U^{\dag}(\pi,0,0), \quad \mathcal{C}_{q}(\rho) = q\mathbb{1}\rho \mathbb{1} + (1-q)U(\pi,0,0)\rho U^{\dag}(\pi,0,0),
\end{equation}
where $\rho$ stands for a $2\times 2$ density matrix. In other words, playing $\mathbb{1}$ and $U(\pi,0,0)$ with probability $p$ and $1-p$ by player 1, and $q$ and $1-q$ by player 2 results also in (\ref{classicalpayoffmixed}). Both ways (\ref{oneparameterd1}) and (\ref{cmstrategies}) turn the EWL game into the classical one. However,  the problem becomes more complex if at least one of the players has access to other unitary operations. The following examples show that the limitation to the probability distributions over the counterparts of classical pure strategies $\mathbb{1}$ and $U(\pi,0,0)$ and considering the EWL game as a $3\times 3$ bimatrix game lose some of the non-classical features of the EWL scheme. 

\begin{example}
\textup{Let us consider the Matching Pennies game in terms of the EWL scheme.  
A common bimatrix form of that game is as follows:}
\begin{equation}\label{matchingpennies}
MP = \bordermatrix{& s^2_{0} & s^2_{1} \cr 
s^1_{0} & (1,-1) & (-1,1) \cr 
s^1_{1} & (-1,1) & (1,-1)}.
\end{equation}
\textup{One can easily show that game (\ref{matchingpennies}) has the unique mixed Nash equilibrium $(\sigma^*_{1}, \sigma^*_{2})$, where $\sigma^*_{1} = [(1/2)(s^1_{0}), (1/2)(s^1_{1})]$ and $\sigma^*_{2} = [(1/2)(s^2_{0}), (1/2)(s^2_{1})]$. 
Let us now extend game (\ref{matchingpennies}) to include the strategy $U(\pi/2,0,-\pi/2)$ for each player. By substituting $\theta_{1} = \theta_{2} = \pi/2$, $\alpha_{1} = \alpha_{2} = 0$ and $\beta_{1} = \beta_{2} = -\pi/2$ into (\ref{payoffvector2x2}) we get }
\begin{equation}
v\left(U\left(\frac{\pi}{2}, 0, -\frac{\pi}{2}\right), U\left(\frac{\pi}{2}, 0, -\frac{\pi}{2}\right)\right) = (0,0).
\end{equation}
\textup{The corresponding bimatrix is of the form}
\begin{equation}\label{qmp}
\bordermatrix{& \mathbb{1} & \mathrm{i}X & U\left(\frac{\pi}{2}, 0, -\frac{\pi}{2}\right) \cr
\mathbb{1} & (1,-1) & (-1,1) & (0,0) \cr
\mathrm{i}X & (-1,1) & (1,-1) & (0,0) \cr
U\left(\frac{\pi}{2}, 0, -\frac{\pi}{2}\right) & (0,0) & (0,0) & (0,0)
}.
\end{equation}
\textup{Among the Nash equilibria are the classical mixed Nash equilibrium} 
\begin{equation}
\left([(1/2)(\mathbb{1}) + (1/2)(\mathrm{i}X)], [(1/2)(\mathbb{1}) + (1/2)(\mathrm{i}X)]\right)
\end{equation}
\textup{and nonclassical Nash equilibria}
\begin{align}
&\left(U\left(\frac{\pi}{2}, 0, -\frac{\pi}{2}\right), U\left(\frac{\pi}{2}, 0, -\frac{\pi}{2}\right)\right),\label{ne1}\\
&\left([(1/2)(\mathbb{1}) + (1/2)(\mathrm{i}X)], U\left(\frac{\pi}{2}, 0, -\frac{\pi}{2}\right)\right), \label{ne2}\\ 
&\left(U\left(\frac{\pi}{2}, 0, -\frac{\pi}{2}\right), [(1/2)(\mathbb{1}) + (1/2)(\mathrm{i}X)]\right). \label{ne3}
\end{align}
\textup{Let us now consider the EWL scheme with unitary strategies}
\begin{equation}\label{newstrategy}
D_{1} = D_{2} = \left\{\{U(\theta,0,0)\colon \theta \in [0,\pi]\} \cup U\left(\frac{\pi}{2}, 0, -\frac{\pi}{2}\right)\right\}.
\end{equation}
\textup{Combining (\ref{payoffvector2x2}) with (\ref{newstrategy}) yields} 
\begin{equation}
v_{1}(U_{1}\otimes U_{2}) = \begin{cases}
\cos{\theta_{1}}\cos{\theta_{2}} &\text{if}~U_{1}\otimes U_{2} = U_{1}(\theta_{1},0,0)\otimes U_{2}(\theta_{2},0,0), \\
-\sin{\theta_{1}} &\text{if}~U_{1}\otimes U_{2} = U_{1}(\theta_{1},0,0)\otimes U_{2}\left(\frac{\pi}{2}, 0, -\frac{\pi}{2}\right), \\ 
-\sin{\theta_{2}} &\text{if}~U_{1}\otimes U_{2} =U_{1}\left(\frac{\pi}{2}, 0, -\frac{\pi}{2}\right)\otimes U_{2}(\theta_{2},0,0), \\
0 &\text{if}~U_{1}\otimes U_{2} = U_{1}\left(\frac{\pi}{2}, 0, -\frac{\pi}{2}\right) \otimes U_{2}\left(\frac{\pi}{2}, 0, -\frac{\pi}{2}\right),
\end{cases}
\end{equation}
\textup{and}
\begin{equation}\label{payoff22}
v_{2}(U_{1}\otimes U_{2}) = -v_{1}(U_{1}\otimes U_{2}).
\end{equation}
\textup{One can show that among  (\ref{ne1}), (\ref{ne2}) and (\ref{ne3}) only strategy profile (\ref{ne2}) is a Nash equilibrium in the game determined by (\ref{newstrategy})-(\ref{payoff22}). In the case of both profiles (\ref{ne1}) and (\ref{ne3}) player 2 obtains the payoff of 0, and she will get the payoff of 1 by choosing $U(\pi/2, 0,0)$,}
\begin{equation}
v_{2}\left(U\left(\frac{\pi}{2}, 0, -\frac{\pi}{2}\right), U\left(\frac{\pi}{2}, 0,0\right)\right) =1. 
\end{equation}
\textup{In general, there is no pure Nash equilibrium in the game given by (\ref{newstrategy})-(\ref{payoff22}). Let us first note that the strategy profile $U(\pi/2, 0,0)\otimes U(\pi/2, 0,0)$ is not a Nash equilibrium. Player 2 can benefit by a unilateral deviation:}
\begin{equation}
1=v_{2}\left(U\left(\frac{\pi}{2}, 0,0\right), U\left(\frac{\pi}{2}, 0,-\frac{\pi}{2}\right)\right) >  v_{2}\left(U\left(\frac{\pi}{2}, 0,0\right), U\left(\frac{\pi}{2}, 0,0\right)\right) = 0.
\end{equation}
\textup{Since there is no other possible Nash equilibria in the set  $\{U_{1}(\theta_{1},0,0)\otimes U_{2}(\theta_{2},0,0)\}$, a strategy profile in the form $U_{1}(\theta_{1},0,0)\otimes U_{2}(\theta_{2},0,0)$ cannot be a Nash equilibrium in the set (\ref{newstrategy}). }

\textup{The last step is to show that neither $U_{1}(\theta_{1},0,0)\otimes U_{2}\left(\frac{\pi}{2}, 0, -\frac{\pi}{2}\right)$ nor $U_{1}\left(\frac{\pi}{2}, 0, -\frac{\pi}{2}\right)\otimes U_{2}(\theta_{2},0,0)$ constitutes a Nash equilibrium. Player 1's best reply to the strategy $U_{2}(\pi/2, 0, -\pi/2)$ is $U_{1}(0,0,0)$ or $U_{1}(\pi,0,0)$ when restricted to the set $\{U_{1}(\theta_{1},0,0)\colon \theta_{1} \in [0,\pi]\}$. But then player 2's best reply to $U_{1}(0,0,0)$ and $U_{1}(\pi,0,0)$ is $U_{2}(\pi,0,0)$ and $U_{2}(0,0,0)$, respectively. Therefore, a strategy profile $U_{1}(\theta_{1},0,0)\otimes U_{2}\left(\frac{\pi}{2}, 0, -\frac{\pi}{2}\right)$ is not a Nash equilibrium. The same conclusion can be drawn for $U_{1}\left(\frac{\pi}{2}, 0, -\frac{\pi}{2}\right)\otimes U_{2}(\theta_{2},0,0)$. This shows that the $3\times3$ bimatrix form used to present the EWL scheme is not equivalent to the original scheme.}
\end{example}
\begin{example}
\noindent \textup{Equally interesting example is the Prisoner's Dilemma game in the form studied in \cite{eisert}:}
\begin{equation}\label{cor}
\bordermatrix{& s^2_{0} & s^2_{1} \cr
s^1_{0} & (3,3) & (0,5) \cr
s^2_{1} & (5,0) & (1,1)}.
\end{equation}
\textup{Let us extend the game in the same manner as (\ref{qmp}). This gives}
\begin{equation}\label{qpd}
\bordermatrix{& \mathbb{1} & \mathrm{i}X & U_{2}\left(\frac{\pi}{2}, 0, -\frac{\pi}{2}\right) \cr
\mathbb{1} & (3,3) & (0,5) & (4,\frac{3}{2}) \cr
\mathrm{i}X & (5,0) & (1,1) & (4,\frac{3}{2}) \cr
U_{1}\left(\frac{\pi}{2}, 0, -\frac{\pi}{2}\right) & (\frac{3}{2},4) & (\frac{3}{2},4) & (\frac{9}{4},\frac{9}{4})
}.
\end{equation}
\textup{Adding $U_{i}\left(\pi/2, 0, -\pi/2\right)$ to the strategy sets of the players in game (\ref{cor}) results in two non-classical equilibria }
\begin{equation}\label{fakenash}
(U_{1}(\pi/2,0, -\pi/2), \mathrm{i}X) \quad \textup{and} \quad (\mathrm{i}X, U_{2}(\pi/2,0, -\pi/2)).
\end{equation}
\textup{Game (\ref{qpd}) is not equivalent to one defined by strategy sets (\ref{newstrategy}). We find that the strategy profiles (\ref{fakenash}) are no longer Nash equilibria in (\ref{newstrategy}). We have}
\begin{equation}
5=v_{1}\left(U_{1}\left(\frac{\pi}{2}, 0,0\right), U_{2}\left(\frac{\pi}{2}, 0, -\frac{\pi}{2}\right)\right) > v_{1}\left(\mathrm{i}X, U_{2}\left(\frac{\pi}{2}, 0, -\frac{\pi}{2}\right)\right)=4
\end{equation}
\textup{and} 
\begin{equation}
5=v_{2}\left(U_{1}\left(\frac{\pi}{2}, 0, -\frac{\pi}{2}\right), U_{2}\left(\frac{\pi}{2}, 0, 0\right)\right) > v_{2}\left(U_{1}\left(\frac{\pi}{2}, 0, -\frac{\pi}{2}\right), \mathrm{i}X\right)=4.
\end{equation}
\end{example}
The above examples demonstrate that adding a single unitary strategy to the bimatrix-form game does not fully reflect nonclassical features of the EWL scheme. The idea of replacing strategy sets of the form $\{U(\theta, 0,0)\}$ with $\{\mathbb{1}, \mathrm{i}X\}$ written with the use of bimatrix form works if strategy set of each player is restricted to the one parameter set. Then a unitary strategy $U(2\arccos{\sqrt{p}},0,0)$ is outcome-equivalent to the mixed strategy $[p(\mathbb{1}), (1-p)(\mathrm{i}X)]$. In general, when other unitary strategies are available the equivalence does not hold. For example, since $v_{1}(\mathbb{1}, U_{2}(\pi/2, 0, \pi/2))$ $=$ $v_{1}(iX, U_{2}(\pi/2, 0, \pi/2))$ $=$ $(a_{00} + a_{10})/2$ for every bimatrix-form game (\ref{bimatrix}), it follows that 
\begin{equation}\label{crownanie}
v_{1}\left([p(\mathbb{1}), (1-p)(\mathrm{i}X)],  U_{2}\left(\frac{\pi}{2}, 0, \frac{\pi}{2}\right)\right) = \frac{a_{00} + a_{10}}{2}.
\end{equation}
In other words, playing any classical mixed strategy against $U_{2}\left(\pi/2, 0, \pi/2\right)$ always results in the same payoff outcome. 
In the case of the strategy profile $\left(U_{1}\left(2\arccos{\sqrt{p}},0,0\right), U_{2}(\pi/2, 0, \pi/2)\right)$, we have
\begin{equation}\label{qrownanie}
v_{1}\left(U_{1}\left(2\arccos{\sqrt{p}},0,0\right), U_{2}\left(\frac{\pi}{2}, 0, \frac{\pi}{2}\right)\right) \\
= \left(\frac{1}{2} + \sqrt{p}\sqrt{1-p}\right)a_{00} + \left(\frac{1}{2}-\sqrt{p}\sqrt{1-p}\right)a_{10}.
\end{equation}
A quick look at Equation~(\ref{qrownanie}) shows the interference terms $\pm\sqrt{p}\sqrt{1-p}$ that are not part of the payoff function (\ref{crownanie}). That is the reason why we obtain different results depending on whether we use strategies of the form $[p(\mathbb{1}), (1-p)(\mathrm{i}X)]$ or the one parameter unitary operations extended with some type of two-parameter operator.
\section{The EWL scheme and the IBM quantum experience}
In what follows, we provide the EWL approach implemented on the IBM quantum experience platform for strategy profiles $(U_{1}(\pi/2, 0, -\pi/2), U_{2}(\pi/2, 0,0))$, $(U_{1}(\pi/2, 0, -\pi/2), \mathbb{1})$ and $(U_{1}(\pi/2, 0, -\pi/2), \mathrm{i}X)$. The quantum circuits are adapted from \cite{nebula}. First, we express unitary operators $U_{1}(\pi/2, 0, -\pi/2)$ and $U_{2}(\pi/2, 0,0))$ in terms of the parametrization of unitary operators used in the IBM quantum circuit composer. Recall that the gates provided by IBM are defined as follows:
\begin{equation}
U^{QC}_{3}(\theta, \phi, \lambda) = \begin{pmatrix} \cos{\frac{\theta}{2}} & -\mathrm{e}^{i\lambda}\sin{\frac{\theta}{2}} \\ 
\mathrm{e}^{i\phi}\sin{\frac{\theta}{2}} & \mathrm{e}^{i(\lambda +\phi)}\sin{\frac{\theta}{2}}
\end{pmatrix},
~~ U^{QC}_{2}(\phi, \lambda) = U^{QC}_{3}\left(\frac{\pi}{2}, \phi, \lambda\right), ~~ U^{QC}_{1}(\lambda) =  U^{QC}_{3}(0,0,\lambda).
\end{equation}
Thus, 
\begin{equation}
U\left(\frac{\pi}{2}, 0, -\frac{\pi}{2}\right) = U^{QC}_{2}(\pi, \pi), ~~ U\left(\frac{\pi}{2}, 0, 0\right) = U_{2}^{QC}\left(\frac{\pi}{2}, -\frac{\pi}{2}\right)
\end{equation}
According to \cite{nebula}, the entangling operator $J$ and the disentangling operator $J^{\dag}$ can be expressed in the form  
\begin{equation}
J = \mathrm{CNOT} \cdot U_{2}(\pi/2, -\pi/2) \cdot \mathrm{CNOT}, ~~ J^{\dag} = \mathrm{CNOT} \cdot U_{2}(-\pi/2, \pi/2) \cdot \mathrm{CNOT}.
\end{equation}
The quantum circuit is presented in Figure~\ref{figure10} (see Appendix for OpenQASM representation of the quantum circuit).
\begin{figure}[t]
\centering\includegraphics[width=5.5in]{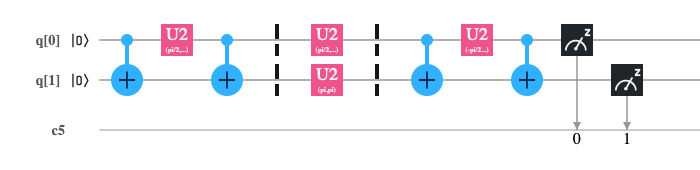}
\caption{Quantum circuit for the EWL scheme. Qubits $q[1]$ and $q[0]$ are identified with the first and second qubit, respectively. Player 1 acts on $q[1]$ with $U^{QC}_{2}(\pi, \pi)$, player 2 acts on $q[0]$ with $U^{QC}_{2}(\pi/2, -\pi/2)$ which corresponds to the strategy profile $(U(\pi/2, 0, -\pi/2), U(\pi/2, 0,0))$ in the EWL approach.}\label{figure10}
\end{figure}

\begin{figure}[t]
\centering\includegraphics[width=5.5in]{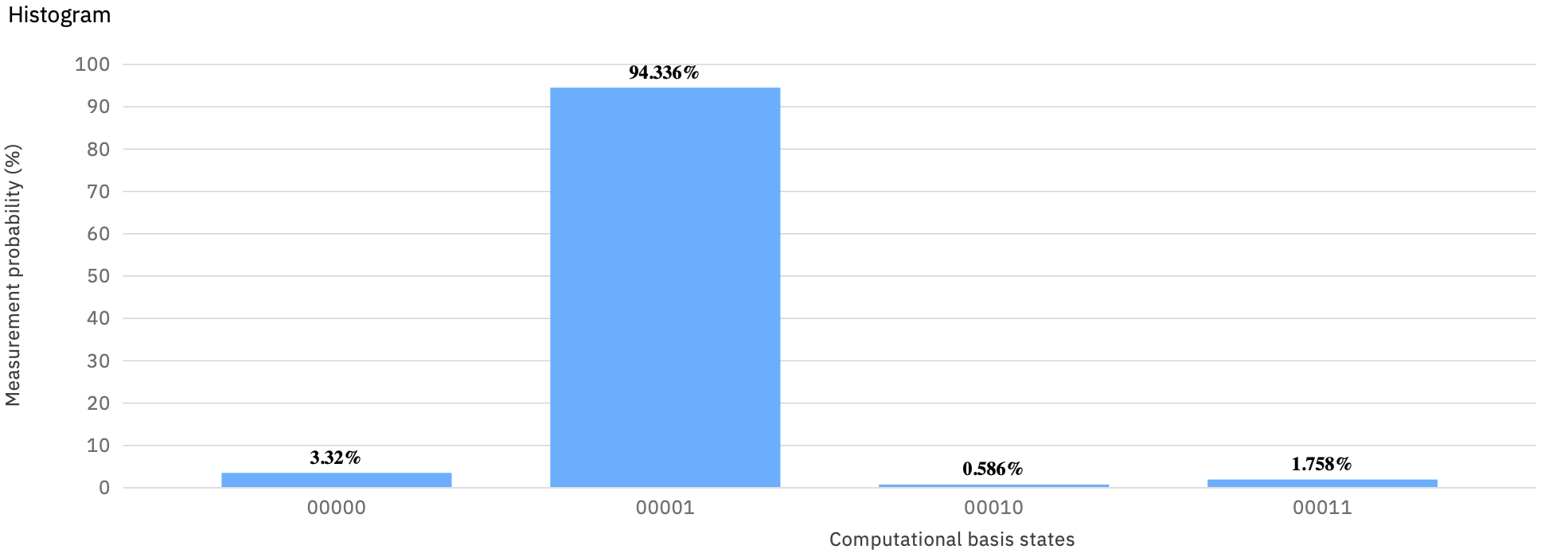}
\caption{The histogram showing the result of the quantum measurement (backend: ibmq\_vigo) corresponding to $(U(\pi/2, 0, -\pi/2), U(\pi/2, 0,0))$ in the EWL approach.}\label{figure12}
\end{figure}

Although, it generates small errors, the IBM quantum computer (ibmq\_vigo) outputs $|01\rangle$ with probability close to one in the case of playing the strategy vector $(U(\pi/2, 0, -\pi/2), U(\pi/2, 0,0))$ or equivalently $(U^{QC}_{2}(\pi, \pi), U^{QC}_{2}(\pi/2, -\pi/2))$ (see Figure~\ref{figure12}). 
\begin{figure}[t]
\centering\includegraphics[width=5.5in]{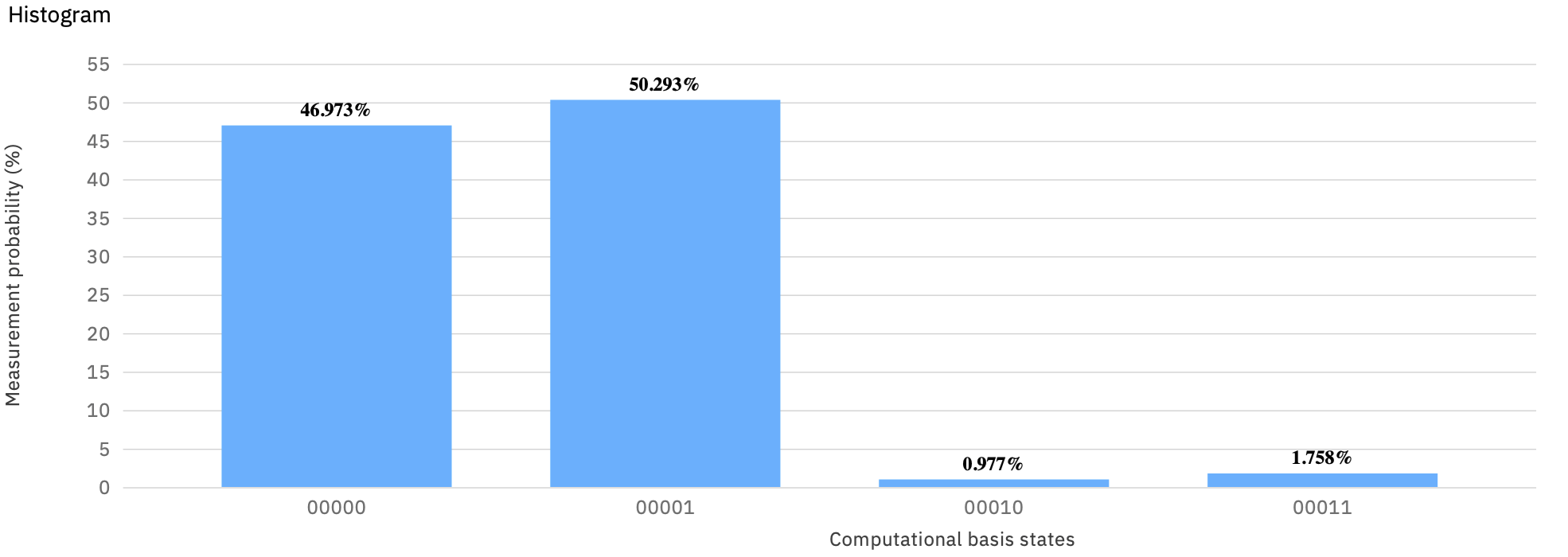}
\caption{The histogram showing the result of the quantum measurement (backend: ibmq\_vigo) corresponding to $(U(\pi/2, 0, -\pi/2), \mathbb{1})$ in the EWL approach.}\label{figure13}
\end{figure}
\begin{figure}[t]
\centering\includegraphics[width=5.8in]{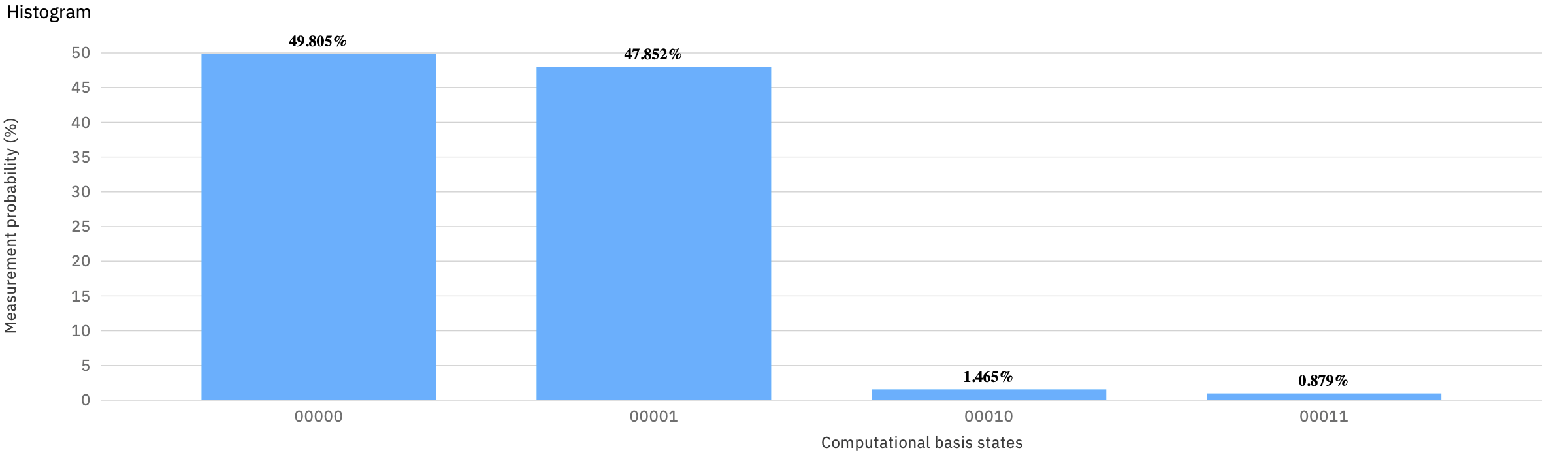}
\caption{The histogram showing the result of the quantum measurement (backend: ibmq\_vigo) corresponding to $(U(\pi/2, 0, -\pi/2), X)$ in the EWL approach.}\label{figure14}
\end{figure}  Comparing the histograms in Figure~\ref{figure12} with ones in Figures \ref{figure13} and \ref{figure14} shows that $U(\pi/2,0,0)$ has no counterpart in any probability distribution over $\mathbb{1}$ and $X$. As a result of playing $\mathbb{1}$ or $X$ against $U(\pi/2, 0, -\pi/2)$ the final state $|\Psi\rangle$  is $|00\rangle$ or $|01\rangle$ with equal probability.   
\section{Payoff region of the EWL quantum game}
\begin{figure}[t]
\centering\includegraphics[width=3.5in]{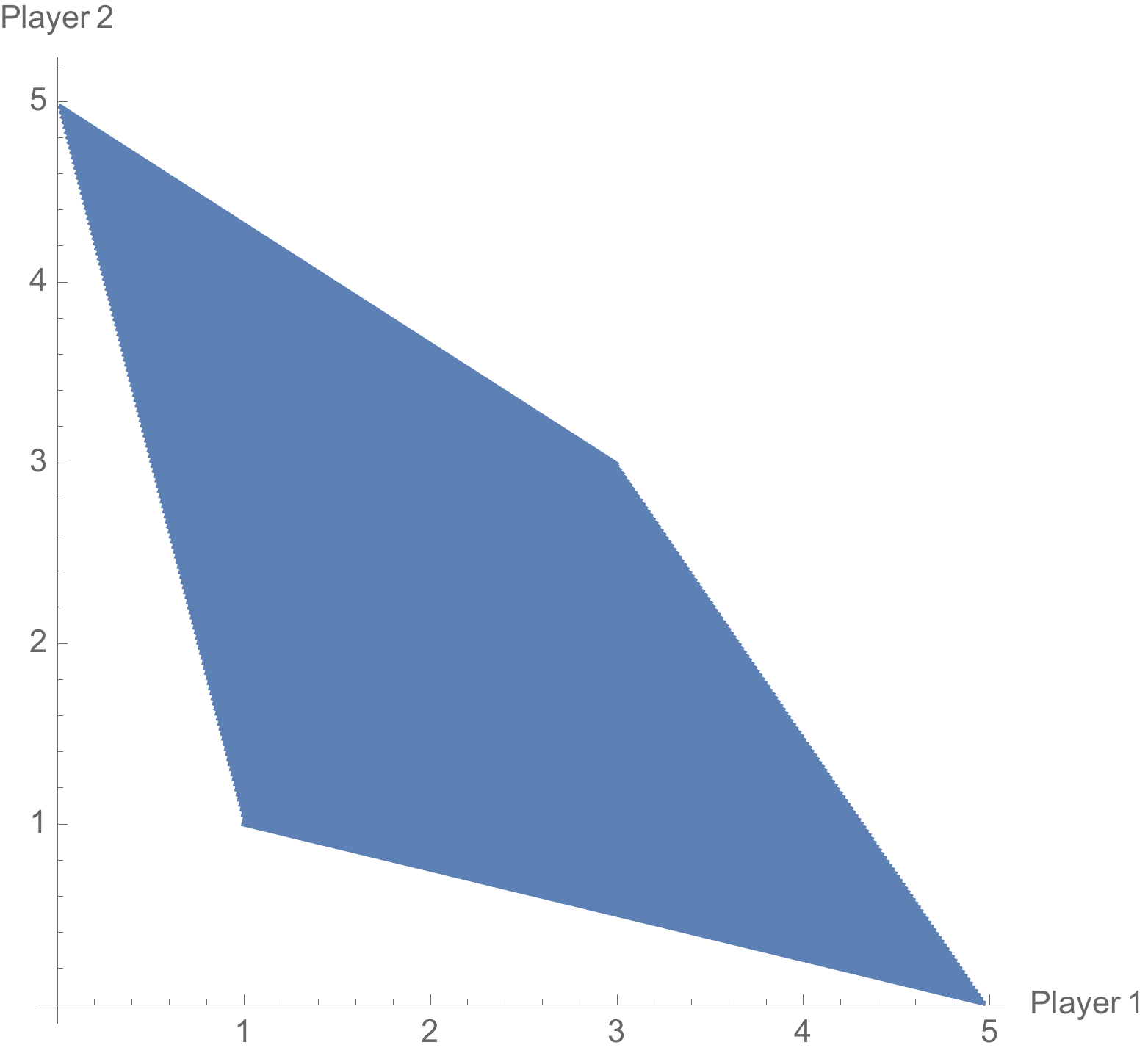}
\caption{Noncooperative payoff region in the Prisoner's Dilemma game that coincides with the cooperative one.}\label{figure1}
\end{figure}

Another advantage that makes the difference between the classical game and the EWL approach is possibility of obtaining payoff profiles which are in the complement of the noncooperative payoff region. The Prisoner's Dilemma game (PD) examined repeatedly with the use of the EWL scheme does not allow one to see that feature. The noncooperative payoff region in the PD game is equal to the cooperative one (see Figure~\ref{figure1}).  
The players by using mixed strategies can obtain each payoff vector from the convex hull of the pure payoff vectors. In general, it is clear that $R_{\textup{pu}} \subsetneq R_{\textup{nc}} \subsetneq R_{\textup{co}}$ (see, Definition~\ref{defR}). The extension of the classical strategies to unitary operators (\ref{2parameter}) makes the sets $R_{\textup{pu}}$, $R_{\textup{nc}}$, $R_{\textup{co}}$ equal in the EWL scheme. The Battle of the Sexes game is a typical example of inequality between the noncooperative and cooperative payoff regions. Its bimatrix form can be written as
\begin{equation}\label{BoS}
BoS = \bordermatrix{& s^2_{0} & s^2_{1} \cr 
s^1_{0} & (4,2) & (0,0) \cr 
s^1_{1} & (0,0) & (2,4)}.
\end{equation}
In this case, the cooperative payoff region is a convex polygon determined by points $(4,2), (2,4)$ and $(0,0)$, and there is no mixed strategy profile from $\Delta(S_{1})\times \Delta(S_{2})$ that would determine the payoff outcome (3,3). The noncooperative and cooperative payoff regions of (\ref{BoS}) are shown in Figure~\ref{figure2}. Mathematica commands for plotting the payoff regions are given in Appendix. 

The outcome (3,3) can be easily achieved by the EWL scheme. From~(\ref{payoffvector2x2}) it follows that 
\begin{equation}
v\left(U_{1}\left(0, \frac{\pi}{8}\right), U_{2}\left(0, \frac{\pi}{8}\right)\right) = \frac{1}{2}((a_{00}, b_{00}) + (a_{11}, b_{11})) = (3,3).
\end{equation}
\begin{figure}[t]
\centering\includegraphics[width=6in]{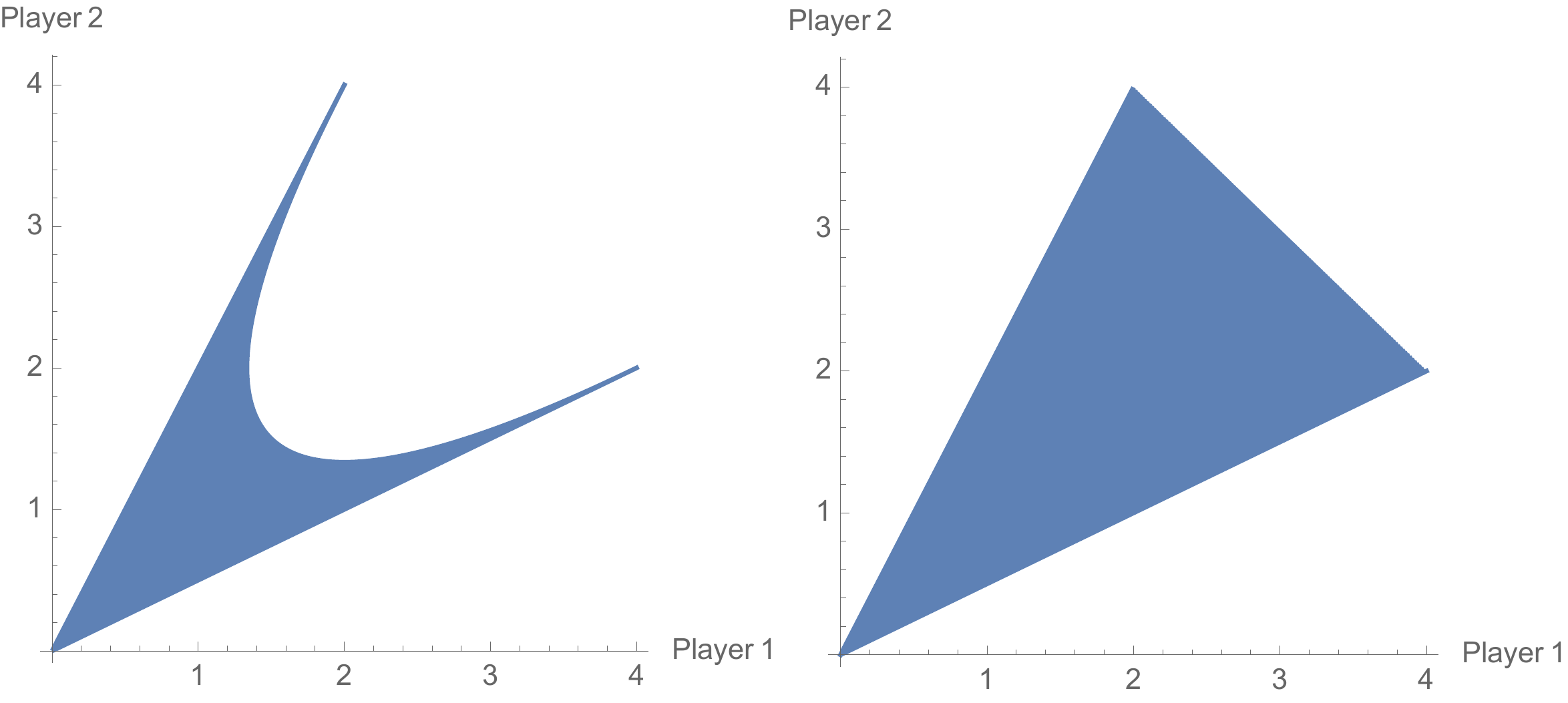}
\caption{Noncooperative payoff region (left) and cooperative payoff region of the Battle of the Sexes game that coincides with the cooperative one.}\label{figure2}
\end{figure}

\noindent In general, the cooperative payoff region of any $2\times 2$ game can be already determined by pure strategy profiles of the two-parameter unitary strategies. We will prove this fact by using the well-known Carath\'eodory's Theorem for convex hulls. 
\begin{theorem}{(Carath\'eodory's Theorem for convex hulls)}\label{carate}
Let A be a subset in $\mathbb{R}^{d}$. Suppose that $x\in \operatorname{conv}(A)$. Then there exists a subset $B$ of $A$ of cardinality at most $n+1$ such that $x\in \operatorname{conv}(B)$.
\end{theorem}
In our case, Carath\'eodory's Theorem states that every payoff vector from $\operatorname{conv}(\{(a_{ij}, b_{ij})\colon i,j=0,1\})$ can be represented as a convex combination of at most three payoff vectors from the pure-payoff region. That observation enables us to prove the following proposition:
\begin{proposition}
The pure payoff region in EWL approach 
\begin{equation}
\Gamma_{EWL} = (\{1,2\}, (U_{i}(\theta_{1}, \alpha_{i}, 0))_{i\in \{1,2\}}, (v_{i})_{i \in \{1,2\}})
\end{equation}
to a general $2\times 2$ game is equal to the cooperative payoff region. 
\end{proposition}
\begin{proof}
It is clear that the pure payoff region of the classical game can be obtained in the EWL game since (\ref{ewlpayofffunction}) coincides with the payoff function of the classical game if the unitary strategies are restricted to the set $\{U(0,0,0), U(\pi,0,0)\}$. 

Let us consider $0 \leq \lambda_{00}, \lambda_{01}, \lambda_{10}, \lambda_{11}  < 1$ such that $\lambda_{00} + \lambda_{01} + \lambda_{10} + \lambda_{11} = 1$. Then there are unitary strategy profiles that depend on $\lambda_{ij}$ and imply a general convex combination of any three pure payoff profiles. Using (\ref{payoffvector2x2}),  we obtain
\begin{align}
&u\left(U_{1}\left(0, -\arccos{\sqrt{\frac{\lambda_{01}}{1-\lambda_{00}}}},0 \right), U_{2}\left(2\arccos{\sqrt{\lambda_{00}}}, \arccos{\sqrt{\frac{\lambda_{01}}{1-\lambda_{00}}}},0\right)\right)\nonumber\\
&\quad =\lambda_{00}(a_{00}, b_{00}) + \lambda_{01}(a_{01}, b_{01}) + \lambda_{10}(a_{10}, b_{10}).\\
&u\left(U_{1}(0,0,0), U_{2}\left(2\arccos{\sqrt{1-\lambda_{01}}}, \arccos{\sqrt{\frac{\lambda_{00}}{1-\lambda_{01}}}},0\right)\right) \nonumber \\
&\quad = \lambda_{00}(a_{00}, b_{00}) + \lambda_{01}(a_{01}, b_{01}) + \lambda_{11}(a_{11}, b_{11}). \\
&u\left(U_{1}\left(0,\frac{\pi}{2},0\right), U_{2}\left(2\arccos{\sqrt{1-\lambda_{10}}}, \arccos{\sqrt{\frac{\lambda_{11}}{1-\lambda_{10}}}},0\right)\right) \nonumber \\ 
&\quad = \lambda_{00}(a_{00}, b_{00}) + \lambda_{10}(a_{10}, b_{10}) + \lambda_{11}(a_{11}, b_{11}).\\
&u\left(U_{1}(\pi,0,0), U_{2}\left(2\arccos{\sqrt{1-\lambda_{11}}}, \arccos{\sqrt{\frac{\lambda_{10}}{1-\lambda_{11}}}},0\right)\right) \nonumber \\ 
&\quad = \lambda_{01}(a_{00}, b_{00}) + \lambda_{10}(a_{10}, b_{10}) + \lambda_{11}(a_{11}, b_{11}).
\end{align}
It follows from Theorem~\ref{carate} that any payoff profile from $\operatorname{conv}(\{(a_{ij}, b_{ij})\colon i,j=0,1\})$ is achievable by the players' pure strategies. In other words, the two-parameter pure strategies in the EWL scheme imply the cooperative payoff region of the corresponding $2\times 2$ game.
\end{proof}
\noindent 
\section{The EWL scheme in relation to van Pike-Enk's arguments}
According to van Enk-Pike comment \cite{enk}, the games written in the form (\ref{qmp}) and (\ref{qpd}) should not be seen as quantum games. They simply describe a $3\times 3$ bimatrix game resulting from the addition of the third pure strategy to the original game. We showed in Section~\ref{mainsec} that bimatrix form cannot fully describe the EWL game since strategies of the form $\{U(\theta,0,0)\mid \theta \in [0,\pi]\}$ are not equivalent to probability distributions over $\mathbb{1}$ and $U(\pi, 0,0)$. As a result, van Pike-Enk's criticism, in fact, does not relate to the original EWL scheme (with continuum of strategies) but merely to a $3\times 3$ bimatrix game with the payoffs calculated by the EWL scheme. 

Still, it was noted in \cite{enk} and \cite{groisman} that adding of another strategy to the classical game changes the rules of the game. Therefore, the outcome resulting from the new game cannot be treated as a solution of the original game. Now, we are going to show that not every extension of strategy sets of the players means changing the rules of the game, in particular, one conducted by unitary strategies in the EWL scheme. A typical example is a mixed extension of the game in which the players can choose probability distributions over their own sets of pure strategies. Let us recall the formal definition of mixed extension of a strategic-form game \cite{maszler}.
\begin{definition}
Let $G = (N, (S_{i})_{i\in N})$ be a strategic-form game (\ref{strgame}) with finite strategy sets. Denote by $S = S_{1}\times S_{2} \times \dots \times S_{n}$ the set of pure strategy vectors. The mixed extension of $G$ is the game
\begin{equation}\label{mixedextension}
\Gamma = (N, (\Sigma_{i})_{i\in N}, (u'_{i})_{i\in N}),
\end{equation}
in which, for each $i \in N$, player $i$'s set of strategies is
\begin{equation}
\Sigma_{i} = \left\{\sigma_{i}\colon S_{i} \to [0,1]\colon \sum_{s_{i} \in S_{i}}\sigma_{i}(s_{i}) = 1\right\}, 
\end{equation}
and her payoff function is the function
\begin{equation}
u'_{i}\colon \Sigma_{1} \times \Sigma_{2} \times \dots \times \Sigma_{n} \to \mathbb{R},
\end{equation}
which associates each strategy vector $\sigma = (\sigma_{1}, \dots, \sigma_{n})$, $\sigma_{i}\in \Sigma_{i}$ with the payoff
\begin{equation}\label{expectedpayoffuprim}
u'_{i}(\sigma) = \sum_{(s_{1}, \dots, s_{n})\in S}u_{i}(s_{1}, \dots, s_{n})\sigma_{1}(s_{1})\sigma_{2}(s_{2})\cdots \sigma_{n}(s_{n}).
\end{equation}
\end{definition}
Nash equilibrium is guaranteed in the mixed extension defined above \cite{nash}. Thus, mixed strategies enable the players to obtain a rational outcome that is not achievable in the set of pure strategy vectors. By using a mixed strategy, a player gets a better payoff in terms of the expected payoff (\ref{expectedpayoffuprim}). Although, it must be assumed that the payoff functions in $G$ satisfy the von Neumann-Morgenstern axioms (see, \cite{maszler}) - their payoff functions are linear in probabilities, it has  nothing to do with breaking the rules of the game $G$. The result of the game $G$ is always a pure strategy vector of $G$. 

Similarly to the mixed extension, the EWL scheme can also be treated as an extension of $G$. The game generated by (\ref{ewlclassic}) is outcome-equivalent to the mixed extension of a $2\times 2$ game if the unitary strategies are restricted to (\ref{oneparameterd1}), and a wider range of unitary operators makes (\ref{ewlclassic}) a nontrivial generalization of (\ref{mixedextension}). Both extensions require using additional resources to be implemented. One would require using some random device to play a mixed strategy. It could be a coin or a dice in the case of simple mixed strategies and a random number generator in general. The unitary strategies, in turn, require using a quantum device.  It is also worth noting that formulas (\ref{ewlpayofffunction}) and (\ref{expectedpayoffuprim}) are just the expected payoff functions. They are associated with specific probability distributions that are generated by the player's mixed strategies and the final state $|\Psi\rangle$. By choosing mixed or unitary strategy, the players create a specific probability distribution over the pure outcomes. However, it is worth emphasising that a mixed extension as well as the EWL approach always result in a pure strategy outcome of $G$.  In the case of the EWL approach to a $2\times 2$ game, the result of the quantum measurement on the final state (determined by the unitary strategies) is one of the four payoff outcomes related to the four pure strategy vectors of the classical game. As stated in \cite{enk}, it would be perfect if the quantum scheme left the classical game unchanged and solved it using quantum operations. In our view, the EWL scheme meets this requirement. 

Mixed and the EWL extensions of a $n$-person strategic-form game (with two-element strategy sets for the players) are summarized in the following table to point out the similarities of two ways of playing the game $G$. 
\begin{center}
\begin{tabular}{|l|}
\hline
Mixed extension $\Gamma = (N, (\Sigma_{i})_{i\in N}, (u'_{i})_{i\in N})$ of a $2\times 2 \times \cdots \times 2$ game \\ \hline $N = \{1,2, \dots, n\}$\\
$\Sigma_{i} = \left\{\sigma_{i}\colon \left\{s^i_{0}, s^i_{1}\right\} \to [0,1]\colon \sigma_{i}(s^i_{0}) + \sigma_{i}(s^i_{1}) = 1\right\}$ \\ 
$u'_{i}(\sigma_{1}, \sigma_{2}, \dots, \sigma_{n}) = \sum_{j_{1}, \dots, j_{n} \in \{0,1\}}u_{i}\left(s^1_{j_{1}}, s^2_{j_{2}}, \dots, s^n_{j_{n}}\right)\sigma_{1}\left(s^1_{j_{1}}\right)\sigma_{2}\left(s^2_{j_{2}}\right)\cdots\sigma_{n}\left(s^n_{j_{n}}\right)$ \\ \hline
\hline
The EWL extension $\Gamma_{EWL} = (N, (D_{i})_{i\in N}, (v_{i})_{i\in N})$ of a $2\times 2 \times \cdots \times 2$ game \\ \hline $N = \{1,2, \dots, n\}$\\
$D_{i} \subset \mathsf{SU}(2) = \left\{U(\theta, \alpha, \beta)\colon \theta \in [0,\pi], \alpha, \beta \in [0, 2\pi)\right\}$ \\ 
$v_{i}(U_{1}, U_{2}, \dots, U_{n}) = \sum_{j_{1}, \dots, j_{n} \in \{0,1\}}u_{i}\left(s^1_{j_{1}}, s^2_{j_{2}}, \dots, s^n_{j_{n}}\right)|\langle \Psi|j_{1}, \dots, j_{n}\rangle|^2$ \\ \hline
\end{tabular}
\end{center}
To sum up, it is not obvious that playing the quantum game really changes the rules of the game if we look at a unitary operator as an extension of a mixed strategy. And if so, it might as well state that using classical mixed strategies violates the rules of the game. The bimatrix games $3\times 3$ in the form of (\ref{qmp}) or (\ref{qpd}) combine outcomes associated with classical pure strategies with one unitary strategy profile determined by the expected payoff function. This way differs significantly from the original scheme presented in \cite{eisert} and cannot be used as an argument against the EWL scheme. 
\section{Conclusions}
The work \cite{eisert} was one of the first papers that launched the quantum game theory. And from that moment on, the idea of  \cite{eisert} has been developed to cover other game theory problems that go beyond simple $2\times 2$ games. The scheme introduced in \cite{eisert} enables the players to obtain the expected payoff outcomes that are often not available when the classical mixed strategies are used. Still, there are doubts if a solution given by the EWL scheme is really of the quantum nature. Among a few comments, it was postulated that the EWL approach to a given game changes the rules of the game. For that reason,  the solution provided by the EWL game should not concern the classical game under study. 

In our opinion, the form of the EWL scheme presented in \cite{eisert} can be regarded as a further generalization of the mixed extension of the game.  In particular case, the EWL approach coincides with the mixed extension since the type of one-parameter unitary operations can be viewed as a counterpart of a mixed strategy. Mixed and the EWL extensions of a game have many features in common that support our view. They both enable the players to obtain a specific probability mixtures of the outcomes and as a result, they generate expected payoff outcomes far beyond the pure payoff region. Noncooperative payoff region is associated with the mixed extension, and the full convex hull of pure payoff vectors (i.e., a cooperative payoff region) is available when the players play the EWL extension of the game. At the same time, the result of the game from playing mixed and unitary strategies is always an outcome from pure payoff region. Another thing is that both extensions have the same structure of strategic-form game. They are both defined by a set of players, sets of players' strategies and the expected payoff functions.  

We think that the EWL scheme does not change the rules of the bimatrix game. As in the case of mixed extension, the EWL extension allows the players to get new possibilities for choosing strategies in the classical game.

\section*{Acknowledgements}
This research was funded by the Pomeranian University in Słupsk. We thank the IBM Quantum team for making   the IBM Quantum Experience.



\section*{Appendix} 
Mathematica commands for plotting the noncooperative payoff region of the Battle of the Sexes game
\vspace{12pt}

\noindent \texttt{A = \{\{4,0\}, \{0,2\}\};}\\
\texttt{B = \{\{2, 0\}, \{0, 4\}\};}\\
\texttt{f[x\_,y\_] = \{x,1-x\}.A.\{y,1-y\};}\\
\texttt{g[x\_,y\_] = \{x,1-x\}.B.\{y,1-y\};}\\
\texttt{h[x\_,y\_] = \{f[x,y], g[x,y]\};} \\
\texttt{table = Table[h[x,y], \{x,0,1,0.002\}, \{y,0,1,0.002\}];}\\
\texttt{flatten = Flatten[table,1];}\\
\texttt{ListPlot[flatten, AspectRatio $\to$ 1, AxesLabel $\to$ \{"Player 1", "Player 2"\}]}
\vspace{12pt}

\noindent Mathematica commands for plotting the payoff region of the EWL approach to the Battle of the Sexes game.
\vspace{12pt}

\noindent \texttt{u1[\textit{t1}\_, \textit{a1}\_, \textit{t2}\_, \textit{a2}\_] = 4$\ast$$\left(\texttt{Cos[a1+a2]}\texttt{Cos}\left[\frac{\texttt{t1}}{\texttt{2}}\right]\texttt{Cos}\left[\frac{\texttt{t2}}{\texttt{2}}\right] \right)^\texttt{2}$} 

 + \texttt{2}$\ast$$\left(\texttt{Sin[a1+a2]}\texttt{Cos}\left[\frac{\texttt{t1}}{\texttt{2}}\right]\texttt{Cos}\left[\frac{\texttt{t2}}{\texttt{2}}\right] - \texttt{Sin}\left[\frac{\texttt{t1}}{\texttt{2}}\right]\texttt{Sin}\left[\frac{\texttt{t2}}{\texttt{2}}\right]\right)^\texttt{2}$\texttt{;}
 
\noindent \texttt{u2[\textit{t1}\_, \textit{a1}\_, \textit{t2}\_, \textit{a2}\_] = 2$\ast$$\left(\texttt{Cos[a1+a2]}\texttt{Cos}\left[\frac{\texttt{t1}}{\texttt{2}}\right]\texttt{Cos}\left[\frac{\texttt{t2}}{\texttt{2}}\right] \right)^\texttt{2}$} 

 + \texttt{4}$\ast$$\left(\texttt{Sin[a1+a2]}\texttt{Cos}\left[\frac{\texttt{t1}}{\texttt{2}}\right]\texttt{Cos}\left[\frac{\texttt{t2}}{\texttt{2}}\right] - \texttt{Sin}\left[\frac{\texttt{t1}}{\texttt{2}}\right]\texttt{Sin}\left[\frac{\texttt{t2}}{\texttt{2}}\right]\right)^\texttt{2}$\texttt{;}

\noindent \texttt{Z[\textit{t1}\_, \textit{a1}\_, \textit{t2}\_, \textit{a2}\_] = \{u1[\textit{t1}\_, \textit{a1}\_, \textit{t2}\_, \textit{a2}\_], u2[\textit{t1}\_, \textit{a1}\_, \textit{t2}\_, \textit{a2}\_]\};}\\
\texttt{tab = Table[Z[\textit{t1}\_, \textit{a1}\_, \textit{t2}\_, \textit{a2}\_], \{t1, 0, $\pi$, 0.06\}, \{t2, 0, $\pi$, 0.1\}}

\texttt{\{a1, 0, 2$\ast\pi$, 0.03\}, \{a2, 0, 2$\ast\pi$, 1\}];}\\
\texttt{Flatten[tab, 3];}\\
\texttt{ListPlot[Flatten[tab, 3], AspectRatio $\to$ 1,} 

\texttt{AxesLabel $\to$ \{"Player 1", "Player 2"\}]}
\vspace{12pt}

\noindent OpenQASM representation of quantum circuit that realizes $(U_{1}(\pi/2, 0, -\pi/2), U_{2}(\pi/2, 0,0))$
\vspace{12pt}

\noindent \texttt{OPENQASM 2.0;}\\
\texttt{include "qelib1.inc";}\\
\texttt{qreg q[15];}\\
\texttt{creg c[5];}\\
\texttt{cx q[0], q[1];}\\
\texttt{u2(1.5707963267948966, -1.5707963267948966) q[0];}\\
\texttt{cx q[0], q[1];}\\
\texttt{u2(1.5707963267948966, -1.5707963267948966) q[0];}\\
\texttt{u2(3.141592653589793, 3.141592653589793) q[1];}\\
\texttt{cx q[0], q[1];}\\
\texttt{u2(-1.5707963267948966, 1.5707963267948966) q[0];}\\
\texttt{cx q[0], q[1];}\\
\texttt{measure q[0] -> c[0];}\\
\texttt{measure q[1] -> c[1];}



\end{document}